\documentclass[12pt,preprint]{aastex}
\usepackage{epsfig}
\usepackage{lscape}

\received{}
\accepted{}
\shorttitle{}
\shortauthors{G\'omez et al.}

\begin{document}

\title{Dynamical decay of a massive multiple system in Orion KL?}

\author{Laura G\'omez, Luis F. Rodr\'\i guez, Laurent Loinard, and
Susana Lizano}
\affil{Centro de Radioastronom\'\i a y Astrof\'\i sica, UNAM,
Apdo. Postal 3-72,\\
Morelia, Michoac\'an, 58089 M\'exico\\
{l.gomez,l.rodriguez,l.loinard,s.lizano@astrosmo.unam.mx}}

\and

\author{Arcadio Poveda and Christine Allen}
\affil{Instituto de Astronom\'\i a, UNAM,
Apdo. Postal 70-264, \\
M\'exico, D. F., 04510 M\'exico\\
{poveda@servidor.unam.mx, chris@astroscu.unam.mx}}

\begin{abstract}
  
  We present absolute astrometry of 35 radio sources in the Orion
  Trapezium and Becklin-Neugebauer/Kleinman-Low regions, obtained from
  Very Large Array archival observations collected over a period of 15
  years. By averaging the results for all the sources, we estimate the
  mean absolute proper motion of Orion to be --in Galactic
  coordinates--   $\mu_\ell \cos b$ = +2.1 $\pm$ 0.2 mas yr$^{-1}$;
  $\mu_b$ = $-$0.1 $\pm$ 0.2 mas yr$^{-1}$. These values agree
  remarkably well with those expected from the differential rotation
  of the Milky Way. Subtraction of this mean motion from the
  individual measurements allows us to register all proper motions to
  the rest frame of the Orion nebula, and identify radio sources with
  large residual velocities. In the KL region, we find three sources
  in this situation:   the BN object, the radio source I, and the
  radio counterpart of the infrared source n. All three objects
  appear to be moving away from a common point where they must all
  have been located about 500 years ago. This suggests that all three
  sources were originally part of a multiple massive stellar system
  that recently disintegrated as a result of a close dynamical
  interaction.

\end{abstract}


\section{Introduction}

Newborn  multiple stars and small stellar clusters tend to be
dynamically unstable. In these systems, close encounters can lead to
the acceleration of one or more of the members, which --when the
acceleration is sufficient-- can escape their birthplace and become
run-away stars. If they affect stars which have not yet completed
their formation, strong dynamical interactions may be expected to
seriously modify the outcome of the star-forming process. If they were
sufficiently common, they could even be partly responsible for the
overall shape of the initial mass function (Reipurth 2000).

As one of the nearest sites of profuse star-formation, Orion is an
obvious candidate   to search for signs of strong dynamical
interactions.  Indeed, one of the most promising examples of a
run-away star is located there: the Becklin-Neugebauer object (Becklin
\& Neugebauer 1967; hereafter BN). The run-away nature of BN was
demonstrated by Plambeck et al.\ (1995) who used multi-epoch radio
observations to show that it moved at a projected velocity of about 50
km s$^{-1}$ toward the north-east, relative to a nearby radio source
taken as reference. Tan (2004) then proposed that it had escaped from
the Trapezium about 4,000 years ago as a result of a dynamical
interaction with the other members of this multiple system. However,
Rodr\'{\i}guez et al.\ (2005) recently showed that another source (the
radio source I --Churchwell et al.\ 1987), located much closer to BN
than the Trapezium, is quickly moving away from BN. Since the
velocities of BN and I are almost exactly anti-parallel,
Rodr\'{\i}guez et al.\ (2005) suggested that these two sources were
originally members of a common system that disintegrated about 500
years ago.

The proper motions reported by Rodr\'{\i}guez et al.\ (2005) were
calibrated against distant quasars, and can therefore be considered
absolute. However, because of the differential rotation of the Galaxy,
Orion is expected to move relatively to these distant quasars, and one
would ideally want to register all motions to the rest-frame of Orion
rather than use absolute values. In the present article, we re-analyze
multi-epoch wide-field Very Large Array (VLA) observations of Orion to
determine the mean absolute motion of Orion. This allows us to show
that yet another source in the KL region   (the radio counterpart of
the infrared source n) can be considered a run-away, and to propose
that the BN object, the radio sources I, and the infrared source n
were all initially members of a common multiple system.

\section{Observations}

The high frequency (22 and 43 GHz) data used by Rodr\'{\i}guez et al.\
(2005) have a limited field of view, and properly sample only the
KL/BN region where a mere handful of radio sources are detected. To
measure the mean absolute proper motion of Orion, one must turn to
lower frequency observations which provide a larger effective field of
view   (several arcminutes), and encompass the Trapezium, BN/KL,
and their surroundings. Thus, we searched the VLA archive for 3.6 and
6 cm observations of Orion taken in the most extended (A)
configuration of the array --the latter requirement ensures that only
data with the highest angular resolution were selected. Four datasets
were retrieved, which span a total of 15 years, with a typical step of
5 years (Tab.\ 1).

  The data were calibrated in AIPS following standard procedures,
and the calibrated visibilities were imaged using weights intermediate
between natural and uniform (with the ROBUST parameter set to 0). To
obtain accurate absolute astrometry, we precessed the data taken in
B1950 to J2000 using the task UVFIX, and used the most recent position
of the phase calibrators for all epochs. With this prescription, we
expect the residual systematic error affecting our data to be at most
of a few milli-arcseconds (mas). Self-calibration was applied to
improve the dynamical range of the images, and allow the detection of
fainter sources. To estimate the effect of the self-calibration
process on our final astrometry, we compared the position of the 15
brightest sources before and after self-calibration. Only small,
random shifts (2-3 mas) were found. These small shifts, as well as the
small possible residual errors due to the uncertainties on the
positions of the phase calibrators mentioned above, were formally
taken into account by adding in quadrature a systematic error of 10
mas to the positional uncertainty delivered by the Gaussian fitting
program (see Sect.\ 3.1 below).

\section{Results}

\subsection{Mean proper motion and velocity dispersion of Orion}

  In total, 35 sources were detected in at least three of our four
epochs (Fig.\ 1a, Tabs. 1 and 2). The position of each source at each
epoch was determined using a linearized least-squares fit to a
Gaussian ellipsoid function (task IMFIT of AIPS).  The final
positional error assigned to each observation is the quadratic sum of
the relative error of the source in the given image (proportional to
the angular resolution over the signal-to-noise ratio) and a
systematic error of 10 mas (see Sect.\ 2). Nearly all of our sources
are very compact ($<$ 0.5$''$ --Tab.\ 2), so we don't expect source
structure to significantly affect our astrometry.

  The source proper motions were then obtained by adjusting their
displacements over the celestial sphere with a linear fit (Tab.\ 3).
To obtain the mean absolute proper motion of Orion, we finally computed
the weighted average of these individual measurements, obtaining:

\begin{center}
$\overline{\mu_\alpha \cos \delta}$ = +0.8 $\pm$ 0.2 mas yr$^{-1}$ 
\vspace{0.2cm}

$\overline{\mu_\delta}$ = $-$2.3 $\pm$ 0.2 mas yr$^{-1}$. 
\end{center}

\noindent
Transformation of these values to the Galactic coordinate system
yields:

\begin{center}
$\overline{\mu_\ell \cos b}$ = +2.1 $\pm$ 0.2 mas yr$^{-1}$ 
\vspace{0.2cm}

$\overline{\mu_b}$ = $-$0.1 $\pm$ 0.2 mas yr$^{-1}$. 
\end{center}

It is interesting to compare these observational values with those
expected from the differential rotation of the Galaxy. The proper
motions determined with the VLA are measured with respect to the
Sun. To obtain the corresponding values expected theoretically, we
will adopt a model for the local rotation of the Galaxy where the Oort
constants are $A$ = 14.4 km s$^{-1}$ kpc$^{-1}$ and $B=-12.0$ km
s$^{-1}$ kpc$^{-1}$ (Allen 2000), and where the distance from the Sun
to the Galactic center is $R_0$ = 8.5 kpc. For the peculiar motion of
the Sun (required to transform the barycentric coordinates provided by
the VLA to values relative to the LSR) , we will use $U_\odot$ = +9.0
km s$^{-1}$, $V_\odot$ = +12.0 km s$^{-1}$, and $W_\odot$ = +7.0 km
s$^{-1}$ (Allen 2000). Here, we follow the traditional convention
where $U$ runs from the Sun to the Galactic center; $V$ is in the
Galactic plane, perpendicular to $U$ and positive in the direction of
Galactic rotation; and $W$ is perpendicular to the Galactic plane,
positive toward the Galactic North pole.   Finally, we will use the
distance estimate to Orion obtained by Genzel et al.\ (1981) applying
the expanding cluster parallax method to a group of H$_2$O masers,
$d$ = 480 $\pm$ 80 pc.  Using these values, and assuming that Orion
is at rest with respect to its LSR, we expect the proper motion of
Orion relative to the Sun to be:

\begin{center}
$\mu_\ell \cos b$ = +1.9 $\pm$ 0.4 mas yr$^{-1}$ 
\vspace{0.2cm}

$\mu_b$ = $-$0.2 $\pm$ 0.2 mas yr$^{-1}$. 
\end{center}

  Here, the error bars account for the uncertainty on the
distance. The agreement between the expected values and the measured
one is --quite remarkably-- better than 0.2 mas yr$^{-1}$, showing
that Orion is indeed nearly at rest with respect to its LSR. An
interesting consequence of this result is that, in spite of its
location nearly 150 pc below the Galactic plane, Orion shows very
little vertical motion relative to it.

  Also using the individual proper motions, we can estimate the
velocity dispersion of the radio sources cluster:

\begin{center}
$\sigma_\alpha$ = 2.3 $\pm$ 0.2 mas yr$^{-1}$ $\equiv$ 5.2 $\pm$ 0.5 km s$^{-1}$ 
\vspace{0.2cm}

$\sigma_\delta$ = 3.1 $\pm$ 0.2 mas yr$^{-1}$ $\equiv$ 7.1 $\pm$ 0.5 km s$^{-1}$. 
\end{center}

The error bars quoted here on the velocity dispersion do not include
the effects of the uncertainty on the distance to Orion --which would
typically contribute an extra 1 km s$^{-1}$ to the errors. The reason
for omitting this contribution is that we shall momentarily compare
the radio and optical velocity dispersions which would be equally
affected by a systematic error on the distance. The values obtained
here for the velocity dispersion are fairly large, and imply a 3-D
velocity dispersion in excess of 10 km s$^{-1}$. Finally, we should
point out that --except in the KL region (see Sect.\ 3.3)-- the
residual velocities do not define an organized pattern (of expansion,
streaming motions or infall), but appear to be random.

\subsection{Comparisons with optical results}

It is interesting to compare the results found here with those
obtained at optical wavelengths. For that purpose, we shall use the
studies of Jones \& Walker (1988 --hereafter JW88) and van Altena et
al.\ (1988 --hereafter vA88). There are about a dozen sources in
common between the catalog of JW88 and the present list of 35 radio
sources, and 3 sources in common between our radio data and the list
of vA88. For the later three sources, the radio and optical
measurements agree very well: to within 1 $\sigma$. We find similarly
good agreement with the sources of JW88 which have modest optical
proper motions. However, there are a few objects for which JW88
measured large proper motions, whereas we find only small
ones. Interestingly, Tian et al. (1996) also noticed that their own
optical proper motion measurements in Orion agreed well with those of
JW88 only for sources with proper motions smaller than 0.6 mas
yr$^{-1}$. Thus, the few sources with large proper motions in the
catalog of JW88 might be less trustworthy than the others, and we
consider that our radio measurements agree overall very well with
published optical ones.

The velocity dispersion for optical measurements is best obtained from
the catalog of JW88, which contains over a thousand sources. Applying
the same weighted average technique used for the radio data, and
restricting ourselves to objects with high membership probability, we
obtain the following velocity dispersions for the optical
measurements:

\begin{center}
$\sigma_\alpha$ = 1.04 $\pm$ 0.02 mas yr$^{-1}$ $\equiv$ 2.37 $\pm$ 0.04 km s$^{-1}$ 
\vspace{0.2cm}

$\sigma_\delta$ = 1.24 $\pm$ 0.02 mas yr$^{-1}$ $\equiv$ 2.83 $\pm$ 0.04 km s$^{-1}$. 
\end{center}

It is noteworthy that the optical dispersion is about 2.5 times
smaller than that obtained with the radio data. This is unlikely to be
a consequence of underestimating the errors associated with the radio
measurements given the remarkable agreement between the average
absolute proper motion of Orion measured at radio wavelengths and the
theoretical expectation (Sect.\ 3.1), and the good agreement between
the radio and optical proper motions for sources where both were
measured. To reinforce this last point, it should be noted that when
there was a significant disagreement between the radio and optical
measurements, the radio observations gave smaller proper motions. If
they played a role, these discrepancies would, therefore, tend to make
the radio velocity dispersion smaller than the optical one, rather
than the opposite.

It is plausible that the radio observations are biased toward a
certain subset of objects with peculiar kinematics. Indeed, it is well
known that low-mass young stars tend to be bright at radio
wavelengths.  Consequently, a large fraction of our sample of radio
sources are likely to be T Tauri stars, whereas the optical
observations will tend to be biased toward brighter (i.e. less
embedded and more massive) stars. The present difference between the
velocity dispersion obtained from optical and radio observations would
then suggest that lower-mass and/or younger sources have larger random
velocities than their older and/or more massive counterparts.

\subsection{Fast-moving sources}

Having measured the mean absolute proper motion of Orion, we are now
in a position to register the motion of all of our sources in the
Orion rest frame. Interestingly, only two sources of the Trapezium
cluster appear to show peculiar residual kinematics: GMR 14 (Garay et
al.\ 1987), and source 46 in the list of Zapata et al.\ (2004). GMR 14
is an extended radio source associated with a proplyd reported by
O'Dell \& Wen (1994). Consequently, the detection of an apparent
motion for this source must be taken with caution, since internal
variability could easily produce changes in the centroid position even
in the absence of a true displacement. As for source 46 in the list of
Zapata et al. (2004), it is highly time variable and located near GMR
23 (at only $0\rlap.{''}6$), which is also a variable radio source. In
such a configuration, differential variability could again easily
produce centroid shifts mimicking position changes, and the present
detection of a large proper motion must be considered very cautiously.
Thus, we consider that the detection of large residual proper motions
for two sources in the Trapezium must be further investigated before
being accepted as fully trustworthy.

The situation is quite different in the BN/KL region, where 3 of the 4
detected sources are found to have residual proper motions above
4$\sigma$ (Fig.\ 2, Tab.\ 3). For sources BN and I, we first confirm
the original finding by Rodr\'{\i}guez et al.\ (2005) that these two
sources are moving away from one another. But the present
determination of the mean absolute proper motion of Orion further
shows that neither source is at rest with respect to Orion. In the
Orion rest frame, BN is moving toward the north-west at 26 km s$^{-1}$
while source I is moving toward the south-east at 15 km s$^{-1}$
(Tab.\ 3).   Since the present proper motions for sources BN and I
are compatible with, but less precise than those reported by
Rodr\'{\i}guez et al. (2005), we shall use the values reported by
these authors --but corrected for the overall motion of Orion measured
here-- in the rest of this paper. In addition to the confirmation of
the large motions of BN and I, we report here for the first time that
the radio counterpart of the infrared source n also has a very large
residual velocity, 24 km s$^{-1}$ approximately toward the south. It
should be pointed out that although n was contained in the field of
view studied by Rodr\'{\i}guez et al.\ (2005), it was not included in
their analysis because it is faint and not detected at high
frequencies. The infrared source n (Lonsdale et al.\ 1982) was
proposed to be a young embedded member either of the KL region, or of
the Trapezium (Lonsdale et al.\ 1982; Wynn-Williams et al.\ 1984;
Dougados et al.\ 1993).

\section{Discussion}

The present results show that there are three sources moving at
projected velocities of 15--25 km s$^{-1}$ within a region only about
10$''$ (0.02 pc) across, centered near the BN object, a situation
which is very unlikely to be coincidental. To investigate the possible
origin of these motions, it is useful to reconstruct the past
positions of the three fast-moving radio sources using their present
locations and velocities --assuming that the latter have remained
constant. Interestingly, it is found that about 500 years ago, all
three sources were --at least in projection-- within a few arcseconds
of each other (Fig.\ 3). Rodriguez et al.\ (2005) had already noticed
that the velocities of BN and I were almost exactly anti-parallel, and
argued that these two sources were originally members of a common
system, that disintegrated about 500 years ago. The present finding
that the double radio source associated with n is also quickly moving
away from the position where Rodr\'{\i}guez et al.\ (2005) had placed
the parental system lends further support to their interpretation, and
suggests that the original multiple system disintegrated in at least
three pieces. An objection which could be made to this interpretation
is the fact that the total momentum of the three sources does not seem
to be 0 when measured in the Orion rest frame. That objection is not
very strong, however, since (i) the parental system could easily have
had a residual motion relative to the Orion rest frame of a few mas
yr$^{-1}$; (ii) the existence of other sources (invisible at
centimeter wavelengths) moving away in different directions cannot be
ruled out. It is also possible that source n is a low-mass object with
a relatively small linear momentum.  In that case, the only two
important contributions are those of BN and I, whose proper motions
average to 0 within the errors.

As discussed at length by Rodr\'{\i}guez et al.\ (2005), in this
disintegration scheme, the current (positive) kinetic energy of the
escaping sources BN, I and n must have been taken from the total
energy of the parental multiple system. Since the latter is assumed to
have been originally bound, its total energy must have been negative.
Conservation of the total energy then dictates that, to compensate for
the excess of   kinetic energy carried by BN, I and n, some components
must have seen their energy become more negative, so they must be more
bound than they originally were. In the classical case of the
disintegration of a non-hierarchical triple system, one of the objects
escapes at high speed, while the other two are re-arranged into a
tight binary. The final total energy of these two bodies is more
negative then it was before the ejection, and the excess of positive
energy thus liberated is carried away by the escaping star. The
simplest generalization to the case of the BN/KL region is that
sources BN, I, and n were initially part of a tight group, which
disintegrated about 500 years ago as a result of a strong dynamical
interaction as in the n-body simulations of Poveda et al.\
(1967). Because of the interaction, BN, I, and n acquired a large total
kinetic energy (about 2 $\times$ 10$^{47}$ ergs if both BN and n are
10 M$_{\odot}$ stars, while source I is a 20 M$_\odot$ star), and one
or more tight binaries were formed.

A related interpretation follows from the recent re-analysis by Bally
\& Zinnecker (2005) of the origin of the massive outflow originating
near sources I and n (Allen \& Burton 1993). This outflow is
associated with gaseous fingers tracing strong bow shocks, and has
traditionally been interpreted as the result of a powerful explosion
(e.g.\ Allen \& Burton 1993; Schultz et al.\ 1999; Bally \& Zinnecker
2005). The analysis of the proper motions of the fingers indicates
that the explosion must have occurred less than a thousand years ago
--a value of 1000 yr is obtained if the velocities have remained
constant, but the actual time elapsed since the explosion may be
somewhat less if there was significant deceleration. Bally \&
Zinnecker (2005) proposed that the explosion may have happened when
source I (a 20 M$_\odot$ star) swallowed a relatively low-mass (1
M$_\odot$) object. Indeed, the total energy liberated by such a merger
is about 3 $\times$ 10$^{48}$ ergs, well in excess of the total energy
carried by the outflow (4 $\times$ 10$^{47}$ ergs --Kwan \& Scoville
1976). Bally \& Zinnecker (2005) also noticed that the epoch of the
explosion coincides roughly with the time when BN and I were very near
to each other, and argued that this was unlikely to be a
coincidence. Consequently, they favored a scenario where the BN object
was ejected from I about 500 years ago, in the same dramatic event
that produced the massive outflow.

The present data suggest that tight binaries have formed as a result
of strong dynamical interactions within a multiple system, but do not
give any information about the ultimate fate of these tight binaries,
i.e.\ if a merger subsequently occurred or not. As an alternative to a
merger, we note that rapid accretion of a disrupted 1 M$_\odot$ disk
around one of the massive stars during the close dynamical interaction
that lead to the decay, would provide a sufficient amount of energy to
power the large H$_2$ outflow. This scenario could explain the near
simultaneity of the dynamical decay and of the explosion that produced 
the flow with no need for a merger.

\section{Conclusions}

In the present paper, we have measured the absolute proper motion of
Orion relative to the Sun using multi-epoch radio observations. The
value we obtain agrees remarkably well with that expected
theoretically from the local rotation of the Milky Way. Using this new
piece of information, we then showed that 3 of the 4 radio sources in
the Orion KL region have large residual velocities. All three sources
appear to move away from a common point of origin, where we argue that
a parental --now defunct-- multiple system must have   been
located. As proposed by Bally \& Zinnecker, the decay of this
original system may be related to the massive H$_2$ flow centered near
Orion KL.

The velocity dispersion of the Orion cluster of radio sources appears
to be nearly three times larger than the velocity dispersion of the 
optical stars, suggesting that there is a systematic difference between
the kinematics of the objects detected at optical and radio wavelengths.
It is plausible that this difference is related to the age or the mass
of the stars preferentially traced by each wavelength. 

\acknowledgments

LFR, LL, and SL are grateful to CONACyT, M\'exico and DGAPA, UNAM for
their support. The National Radio Astronomy Observatory (NRAO) is a
facility of the National Science Foundation operated under cooperative
agreement by Associated Universities, Inc.

\begin{deluxetable}{lccccccc}
  \tablewidth{17.0cm} \tablecaption{Archive Observations Analyzed}
  \tablehead{ \colhead{} & \colhead{} & \colhead{t$_{int}$} & \colhead{$\lambda$} & \colhead{$\Delta \nu$\tablenotemark{b}}& \colhead{Phase}
    & \colhead{Synthesized Beam} & \colhead{$\sigma$}\\
    \colhead{Epoch} & \colhead{VLA code} & \colhead{(hrs)}& \colhead{(cm)} & \colhead{MHz} &\colhead{Calibrator}
    & \colhead{($\theta_{max} \times \theta_{min}; PA$)\tablenotemark{a}} & \colhead{($\mu$Jy beam$^{-1}$)}\\
  } \startdata
  1985.05 & AM143 & 0.6 & 6.0 & 100 & 0541-056 & $0\rlap.{''}43 \times 0\rlap.{''}35;~-15^\circ$ & 136\\
  1991.68 & AM335 & 1.8 & 3.6 & 100 & 0501-019\tablenotemark{c} & $0\rlap.{''}26 \times 0\rlap.{''}25;~-55^\circ$ & 77\\
  1995.56 & AM494 & 8.2 & 3.6 & 31.2 & 0541-056 & $0\rlap.{''}26 \times 0\rlap.{''}22;~+34^\circ$ & 42\\
  2000.87 & AM668 & 9.4 & 3.6 & 31.2 &0541-056 & $0\rlap.{''}24 \times 0\rlap.{''}22;~+3^\circ$ & 40\\
  \enddata \tablenotetext{a}{Major axis$\times$minor axis in arcsec;
    PA in degrees.} \tablenotetext{b} {Effective bandwidth for both circular 
polarization, combining the two IFs.}
   \tablenotetext{c} {For these observations the
    source 0530+135 was also used as phase calibrator.}
\end{deluxetable}

\begin{deluxetable}{llcccccccr}
 \tablewidth{16cm}
 \tablecaption{General properties of the radio sources\tablenotemark{a}}
  \tablehead{ 

\colhead{Source} & 
\colhead{Other} &
\multicolumn{3}{c}{$\alpha_{J2000.0}$} &
\multicolumn{3}{c}{$\delta_{J2000.0}$}&
\colhead{Flux density} &
\colhead{Size} \\%

\colhead{\#} &
\colhead{name} &
\colhead{($^h$)} &
\colhead{($^m$)} &
\colhead{($^s$)} &
\colhead{($^\circ$)} &
\colhead{($^{'}$)} &
\colhead{($^{''}$)} & 
\colhead{(mJy)} & 
\colhead{($''$)} \\%
  } \startdata
1  & GMR A     & 05 & 35 & 11.8022   & $-$05 & 21 & 49.229  & 12.2 &  $<$0.1 \\%
2  & BN object & 05 & 35 & 14.1131   & $-$05 & 22 & 22.793  &  3.8 &  $<$0.1 \\%
3  & GMR C     & 05 & 35 & 14.1614   & $-$05 & 23 & 01.129  &  6.7 &   0.7 \\%
4  & IR n      & 05 & 35 & 14.3553   & $-$05 & 22 & 32.702  &  1.6 &   0.6 \\%
5  & GMR I     & 05 & 35 & 14.5121   & $-$05 & 22 & 30.521  &  0.7 &  $<$0.1 \\%
6  & GMR D     & 05 & 35 & 14.8969   & $-$05 & 22 & 25.394  &  4.1 &  $<$0.1 \\%
7  & GMR 14    & 05 & 35 & 15.5226   & $-$05 & 23 & 37.375  &  4.9 &   0.4 \\%
8  & GMR 26    & 05 & 35 & 15.7288   & $-$05 & 23 & 22.477  &  3.1 &   0.2 \\%
9  & GMR 13    & 05 & 35 & 15.7964   & $-$05 & 23 & 26.562  & 11.3 &   0.3 \\%
10 & GMR 12    & 05 & 35 & 15.8243   & $-$05 & 23 & 14.123  & 28.1 &  $<$0.1 \\%
11 & GMR 11    & 05 & 35 & 15.8393   & $-$05 & 23 & 22.480  & 11.5 &   0.2 \\%
12 & GMR 10    & 05 & 35 & 15.8488   & $-$05 & 23 & 25.540  &  6.2 &   0.3 \\%
13 & GMR 24    & 05 & 35 & 15.9015   & $-$05 & 23 & 37.970  &  2.3 &  $<$0.1 \\%
14 & GMR 9     & 05 & 35 & 15.9508   & $-$05 & 23 & 49.801  &  9.9 &   0.5 \\%
15 & Zapata 46 & 05 & 35 & 15.9971   & $-$05 & 23 & 52.940  &  2.3 &   0.3 \\%
16 & GMR 8     & 05 & 35 & 16.0674   & $-$05 & 23 & 24.333  &  5.7 &  $<$0.1 \\%
17 & GMR 15    & 05 & 35 & 16.0716   & $-$05 & 23 & 07.073  &  5.1 &   0.2 \\%
18 & GMR 22    & 05 & 35 & 16.0776   & $-$05 & 23 & 27.826  &  1.9 &  $<$0.1 \\%
19 & GMR 7     & 05 & 35 & 16.2890   & $-$05 & 23 & 16.575  & 10.8 &   0.2 \\%
20 & GMR 16    & 05 & 35 & 16.3269   & $-$05 & 23 & 22.597  &  3.7 &  $<$0.1 \\%
21 & GMR K     & 05 & 35 & 16.3986   & $-$05 & 22 & 35.315  &  1.4 &   0.2 \\%
22 & GMR 21    & 05 & 35 & 16.6190   & $-$05 & 23 & 16.096  &  1.9 &  $<$0.1 \\%
23 & GMR 6     & 05 & 35 & 16.7527   & $-$05 & 23 & 16.452  & 25.0 &   0.2 \\%
24 & GMR 17    & 05 & 35 & 16.7694   & $-$05 & 23 & 28.036  &  3.9 &  $<$0.1 \\%
25 & GMR 5     & 05 & 35 & 16.8466   & $-$05 & 23 & 26.202  & 18.5 &   0.4 \\%
26 & GMR E     & 05 & 35 & 16.9716   & $-$05 & 22 & 48.677  &  2.8 &   0.4 \\%
27 & GMR 4     & 05 & 35 & 16.9796   & $-$05 & 23 & 36.984  &  9.5 &   0.3 \\%
28 & GMR 3     & 05 & 35 & 17.0665   & $-$05 & 23 & 34.027  &  4.3 &  $<$0.1 \\%
29 & GMR L     & 05 & 35 & 17.3514   & $-$05 & 22 & 35.897  &  2.4 &   0.4 \\%
30 & GMR 2     & 05 & 35 & 17.5605   & $-$05 & 23 & 24.863  &  5.0 &   0.2 \\%
31 & GMR 1     & 05 & 35 & 17.6739   & $-$05 & 23 & 40.908  &  9.7 &   0.5 \\%
32 & GMR G     & 05 & 35 & 17.9489   & $-$05 & 22 & 45.468  &  3.4 &  $<$0.1 \\%
33 & GMR 19    & 05 & 35 & 18.0447   & $-$05 & 23 & 30.719  &  4.6 &   0.2 \\%
34 & Zapata 75 & 05 & 35 & 18.2422   & $-$05 & 23 & 15.617  &  0.7 &  $<$0.1 \\%
35 & GMR F     & 05 & 35 & 18.3706   & $-$05 & 22 & 37.436  &  3.0 &  $<$0.1 \\%
  \enddata 
\tablenotetext{a}{The positions, flux densities and sizes reported here are from 
the 3.6 cm November 13, 2000 observation.}
\end{deluxetable}

\begin{deluxetable}{llrrrr}
  \tablewidth{15.5cm}
\tablecaption{Proper Motions of the radio sources measured in the Orion rest frame\tablenotemark{a}}
  \tablehead{ 
\colhead{Source} & 
\colhead{Other} &
\colhead{$\mu_{\alpha} \cos \delta$} &
\colhead{$\mu_{\delta}$} &
\colhead{$\mu_{total}$} &
\colhead{P.A.} \\%
\colhead{\#} &
\colhead{name} &
\colhead{(mas yr$^{-1}$)} & 
\colhead{(mas yr$^{-1}$)} & 
\colhead{(mas yr$^{-1}$)} & 
\colhead{($^\circ$)} \\%
  } \startdata
1  & GMR A     & $-$2.82  $\pm$ 1.86 &     3.45  $\pm$ 1.92  &  4.45  $\pm$ 1.90  &   $-$39.2 $\pm$ 24.4  \\%
2  & BN object & $-$5.90  $\pm$ 1.28 &     9.45  $\pm$ 1.37  & 11.14  $\pm$ 1.35  &   $-$32.0 $\pm$  6.9  \\%
3  & GMR C     & $-$0.95  $\pm$ 1.01 &  $-$3.34  $\pm$ 1.24  &  3.47  $\pm$ 1.22  &  $-$164.2 $\pm$ 20.2  \\%
4  & IR n\tablenotemark{b}      &    1.78  $\pm$ 2.10 & $-$11.72  $\pm$ 2.20  & 11.85  $\pm$ 2.20  &     171.3 $\pm$ 10.6  \\%
5  & GMR I     &    2.26  $\pm$ 1.92 &  $-$7.46  $\pm$ 1.89  &  7.79  $\pm$ 1.89  &     163.1 $\pm$ 13.9  \\%
6  & GMR D     & $-$0.41  $\pm$ 1.57 &  $-$4.62  $\pm$ 1.56  &  4.64  $\pm$ 1.56  &  $-$175.0 $\pm$ 19.3  \\%
7  & GMR 14    &    5.05  $\pm$ 1.05 &     5.37  $\pm$ 1.06  &  7.38  $\pm$ 1.06  &      43.3 $\pm$  8.2  \\%
8  & GMR 26    &    0.84  $\pm$ 1.14 &  $-$1.45  $\pm$ 1.24  &  1.68  $\pm$ 1.22  &     149.8 $\pm$ 41.6  \\%
9  & GMR 13    &    1.71  $\pm$ 0.92 &     2.83  $\pm$ 0.92  &  3.31  $\pm$ 0.92  &      31.2 $\pm$ 15.9  \\%
10 & GMR 12    &    3.09  $\pm$ 1.54 &  $-$1.98  $\pm$ 1.53  &  3.67  $\pm$ 1.54  &     122.6 $\pm$ 24.0  \\%
11 & GMR 11    &    0.96  $\pm$ 0.89 &  $-$0.09  $\pm$ 0.89  &  0.97  $\pm$ 0.89  &      95.2 $\pm$ 52.7  \\%
12 & GMR 10    &    0.61  $\pm$ 0.98 &  $-$3.03  $\pm$ 1.01  &  3.09  $\pm$ 1.01  &     168.5 $\pm$ 18.7  \\%
13 & GMR 24    & $-$0.14  $\pm$ 1.64 &  $-$1.82  $\pm$ 1.64  &  1.82  $\pm$ 1.64  &  $-$175.7 $\pm$ 51.5  \\%
14 & GMR 9     & $-$2.89  $\pm$ 1.01 &     1.06  $\pm$ 1.01  &  3.08  $\pm$ 1.01  &   $-$69.8 $\pm$ 18.8  \\%
15 & Zapata 46 & $-$5.66  $\pm$ 2.37 &     8.64  $\pm$ 2.12  & 10.33  $\pm$ 2.20  &   $-$33.2 $\pm$ 12.2  \\%
16 & GMR 8     &    0.53  $\pm$ 0.97 &  $-$3.09  $\pm$ 0.97  &  3.13  $\pm$ 0.97  &     170.2 $\pm$ 17.7  \\%
17 & GMR 15    & $-$0.49  $\pm$ 0.97 &     0.07  $\pm$ 0.98  &  0.49  $\pm$ 0.97  &   $-$81.6 $\pm$ 113.2 \\%
18 & GMR 22    &    2.67  $\pm$ 1.14 &     1.89  $\pm$ 1.10  &  3.28  $\pm$ 1.13  &      54.7 $\pm$ 19.7  \\%
19 & GMR 7     &    0.06  $\pm$ 0.90 &     0.48  $\pm$ 0.91  &  0.49  $\pm$ 0.91  &       7.6 $\pm$ 107.2 \\%
20 & GMR 16    & $-$1.41  $\pm$ 1.01 &  $-$0.52  $\pm$ 1.04  &  1.50  $\pm$ 1.01  &  $-$110.2 $\pm$ 38.8  \\%
21 & GMR K     & $-$4.43  $\pm$ 2.06 &  $-$3.05  $\pm$ 2.00  &  5.37  $\pm$ 2.04  &  $-$124.5 $\pm$ 21.8  \\%
22 & GMR 21    &    0.00  $\pm$ 1.32 &  $-$6.39  $\pm$ 1.26  &  6.39  $\pm$ 1.26  &     180.0 $\pm$ 11.3  \\%
23 & GMR 6     &    1.20  $\pm$ 0.88 &  $-$1.92  $\pm$ 0.88  &  2.26  $\pm$ 0.88  &     147.9 $\pm$ 22.3  \\%
24 & GMR 17    & $-$2.48  $\pm$ 1.03 &  $-$1.95  $\pm$ 1.02  &  3.15  $\pm$ 1.03  &  $-$128.2 $\pm$ 18.7  \\%
25 & GMR 5     & $-$1.81  $\pm$ 0.89 &  $-$1.30  $\pm$ 0.91  &  2.22  $\pm$ 0.90  &  $-$125.7 $\pm$ 23.1  \\%
26 & GMR E     & $-$2.50  $\pm$ 1.24 &  $-$0.90  $\pm$ 1.14  &  2.65  $\pm$ 1.23  &  $-$109.8 $\pm$ 26.5  \\%
27 & GMR 4     &    0.46  $\pm$ 0.92 &     3.53  $\pm$ 0.93  &  3.56  $\pm$ 0.93  &       7.5 $\pm$ 15.0  \\%
28 & GMR 3     &    2.13  $\pm$ 0.98 &     2.33  $\pm$ 0.98  &  3.16  $\pm$ 0.98  &      42.5 $\pm$ 17.8  \\%
29 & GMR L     &    0.80  $\pm$ 1.64 &     3.18  $\pm$ 1.32  &  3.28  $\pm$ 1.34  &      14.2 $\pm$ 23.4  \\%
30 & GMR 2     & $-$3.12  $\pm$ 0.98 &     0.85  $\pm$ 0.96  &  3.23  $\pm$ 0.98  &   $-$74.7 $\pm$ 17.4  \\%
31 & GMR 1     & $-$2.98  $\pm$ 0.95 &     0.27  $\pm$ 0.98  &  2.99  $\pm$ 0.95  &   $-$84.8 $\pm$ 18.2  \\%
32 & GMR G     &    1.08  $\pm$ 1.16 &     3.01  $\pm$ 1.21  &  3.20  $\pm$ 1.20  &      19.8 $\pm$ 21.6  \\%
33 & GMR 19    & $-$0.25  $\pm$ 1.01 &  $-$1.88  $\pm$ 1.00  &  1.89  $\pm$ 1.00  &  $-$172.5 $\pm$ 30.3  \\%
34 & Zapata 75 &    5.62  $\pm$ 2.33 &  $-$0.56  $\pm$ 2.05  &  5.65  $\pm$ 2.33  &      95.7 $\pm$ 23.6  \\%
35 & GMR F     &    3.58  $\pm$ 0.87 &     0.30  $\pm$ 0.87  &  3.60  $\pm$ 0.87  &      85.2 $\pm$ 13.9  \\%
  \enddata 
\tablenotetext{a}{The errors quoted in this Table are 1--$\sigma$.}
\tablenotetext{b}{The proper motions reported here were obtained by
averaging the motion of the two components of this double radio source.}
\end{deluxetable}

\begin{figure*}
\centerline{\includegraphics[height=1.00\textwidth,angle=-90]{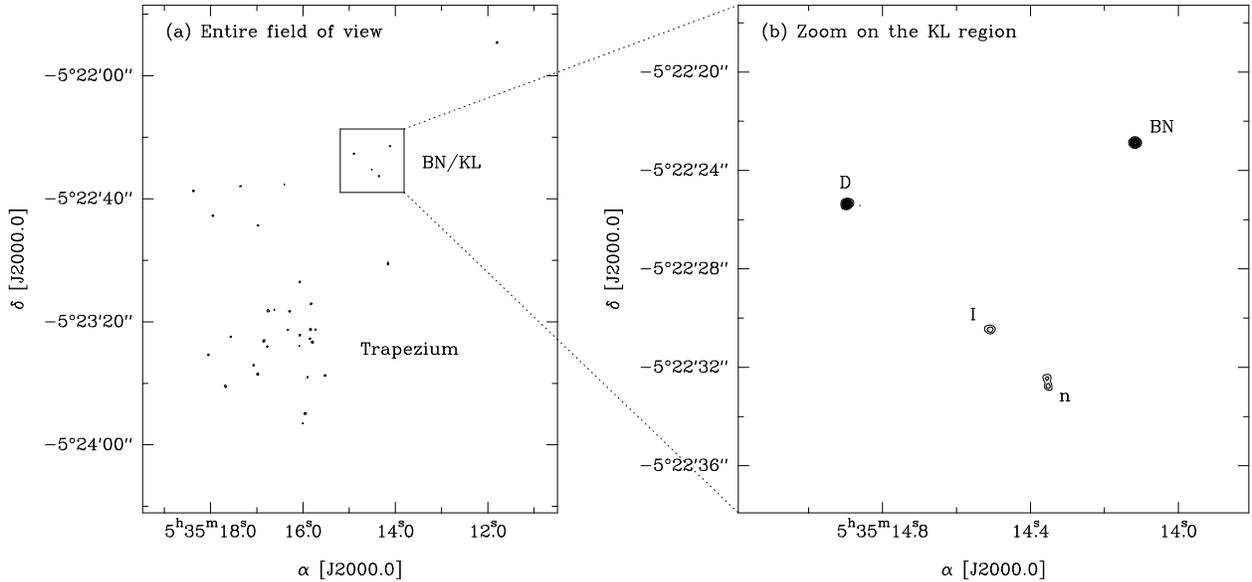}}
\caption{3.6 cm VLA images at epoch 1991.68. (a) Image of the
  entire field of view studied. The only contour shown is at 0.8 mJy
  beam$^{-1}$. Over 30 sources are visible in that image. (b) Zoom on
  the KL region. The first contour and the contour interval are at 0.4
  mJy beam$^{-1}$. The half power contour of the synthesized beam
  is $0\rlap.{''}26 \times 0\rlap.{''}25$; PA = $-55^\circ$.}
\end{figure*}

\begin{figure}
\centerline{\includegraphics[height=0.8\textwidth,angle=0]{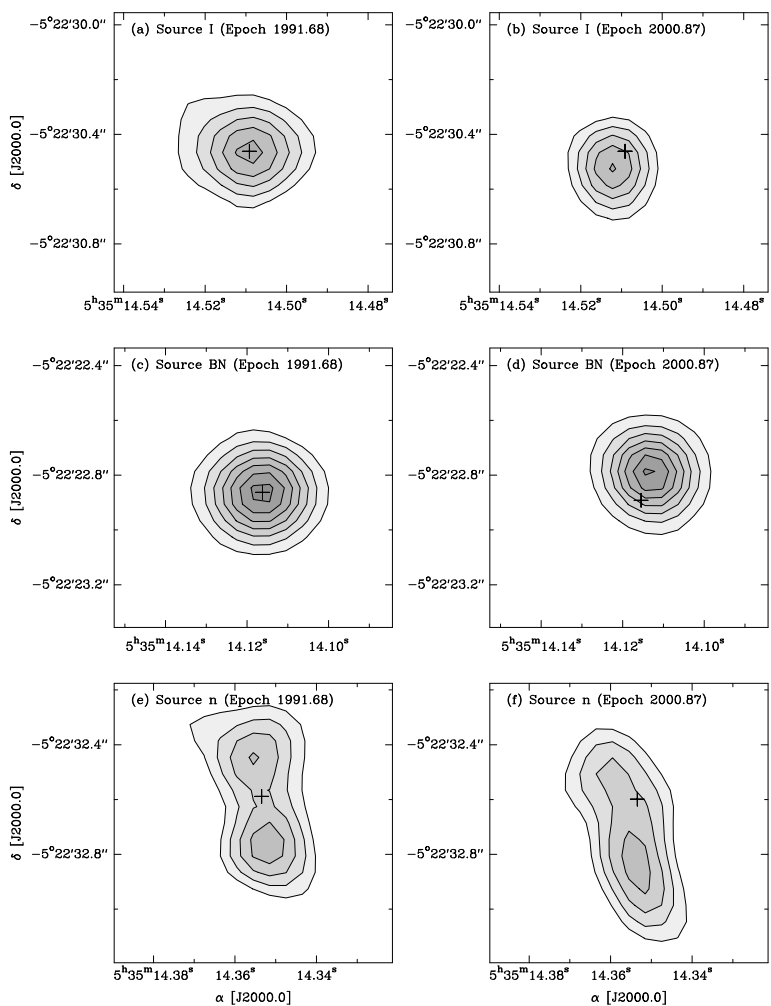}}
\caption{Comparison between the positions of the sources BN, I and n
at the first and last of our observations. In each row, the cross
shows the center position of the sources at the first epoch. The first
contours and the contour spacings are 0.2 mJy beam$^{-1}$ for source I
and n, and 0.5 mJy beam$^{-1}$ for BN.}
\end{figure}

\begin{figure}
\centerline{\includegraphics[height=0.5\textwidth,angle=-90]{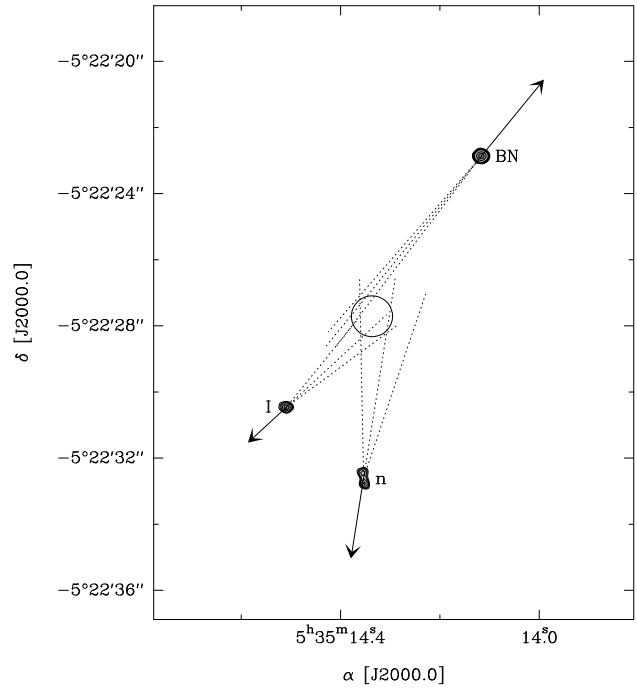}}
\caption{Diagram showing the proper motions of the sources in the BN/KL 
  region. Here we used the better determined proper motions provided
  by Rodr\'{\i}guez et al.\ (2005) for sources BN and I, and the value
  found in the present paper for source n . The arrows show the
  direction and amplitude of the source velocities, and the dotted
  lines encompass their past positions. About 500 years ago, all three
  sources must have been located in the small circle, within a few
  arcseconds of each other.}
\end{figure}

\end{document}